\documentclass[a4paper,fleqn]{cas-sc}

\usepackage[authoryear]{natbib}
\usepackage{graphicx} 
\usepackage{float}
\usepackage{algorithm}  
\usepackage{algpseudocode}
\usepackage{color}
\usepackage{setspace}
\usepackage[nomarkers,figuresonly]{endfloat}
\usepackage{soul}

\def\tsc#1{\csdef{#1}{\textsc{\lowercase{#1}}\xspace}}
\tsc{WGM}
\tsc{QE}
\tsc{EP}
\tsc{PMS}
\tsc{BEC}
\tsc{DE}
\usepackage{lineno}
\begin{document}
\let\WriteBookmarks\relax
\def\floatpagepagefraction{1}
\def\textpagefraction{.001}
\shorttitle{Crust Macrofracturing as Evidence of Last Deglaciation}
\shortauthors{Aleshin et al.}

\title [mode = title]{Crust Macrofracturing as the Evidence of the Last Deglaciation
}

\author[1]{I.M.~Aleshin}
[type=editor,auid=000,bioid=1,orcid=0000-0001-7489-2951]
\credit{Supervision, Formal Analysis, Conceptualization, Methodology, Project administration, Software, Writing - original draft, Writing - review \& editing}

\author[1]{K.I.~Kholodkov}
[orcid=0000-0003-0324-9795]
\credit{Software, Visualization, Writing - review \& editing}

\author[2]{E.G.~Kozlovskaya}
\credit{Validation, Data curation, Methodology, Writing - original draft,Writing - review \& editing}

\author[1]{I.V.~Malygin}
\credit{Software, Formal Analysis}

\address[1]{Schmidt Institute of Physics of the Earth of the Russian Academy of Sciences, Moscow, Russia}
\address[2]{Oulu, Finland} 

\begin{abstract}
Machine learning methods were applied to reconsider the results of several passive seismic experiments in Finland. 
We created datasets from different stages of the receiver function technique and processed them with one of basic machine learning algorithms. 
All the results were obtained uniformly with the k-nearest neighbors algorithm.
The first result is the Moho depth map of the region. 
Another result is the delineation of the near-surface low $S$-wave velocity layer. 
There are three such areas in the Northern, Southern, and central parts of the region. 
The low $S$-wave velocity in the Northern and Southern areas can be linked to the geological structure. 
However, we attribute the central low $S$-wave velocity area to a large number of water-saturated cracks in the upper 1-5 km. 
Analysis of the structure of this area leads us to the conclusion that macrofracturing was caused by the last deglaciation. 
\end{abstract}
 

\maketitle 

\printcredits

\doublespacing

\section{Introduction}
\label{intro}

Passive seismic experiments SVEKALAPKO \citep{bock} and POLENET/LAPNET \citep{kozlovskaya2006} spurred a large number of research activities that targeted the lithosphere structure of the northern part of the Baltic Shield \citep{bruneton2002, alinaghi2003, aleshin2006, vecsey2007, kozlovskaya2008, uski2011, grad2012, silvennoinen2014, VINNIK2016}. 
A significant part of these studies is based on the receiver function technique \citep{alinaghi2003, aleshin2006, kozlovskaya2008, grad2012, VINNIK2016}. 
With data from a wide network of seismic stations across the region, this method yielded several important results on the structure of the mantle of northern and southern Finland \citep{alinaghi2003, frassetto2013, VINNIK2016}, and played a key role in the implementation of a three-dimensional seismic model of central and southern Finland \citep{kozlovskaya2008}.
An important structural feature of the region is the presence of a thin surface layer with low shear wave velocities Vs. 
The layer thickness varies around 1 km, the Vs values are down to 40 percent less than the average over the crust \citep{aleshin2006}. 
The presence of this layer was first noted in \citep{pedersen1991,grad1992,grad1994} in the study of the seismic surface wave attenuation. 
They reveal not only low values of $S$-wave velocity in the shallow crust but also low values of seismic quality factor (Q-factor), which increase rapidly down to the depth of about 1 km. According to \cite{grad1998}, the Q-factor is suitable to differentiate the Archean and the Proterozoic crust of the Fennoscandian Shield. In \cite{grad1992} showed that the increased density of cracks provide for lower Q-factor for the top 1 km of the crust. This explains the low $S$-wave velocity.
The presence of the low $S$-wave velocity surface layer (LVSL) was confirmed later in \cite{aleshin2006}. However, \cite{kozlovskaya2008} have shown that the presence of such a layer is independent of the age of the rocks, it is present in a significant part of both Archean and Proterozoic regions. 
The research of the Moho shape adopts the results of our previous studies in the SVEKALAPKO and POLENET/LAPNET projects \citep{aleshin2006,kozlovskaya2008,silvennoinen2014} supplemented by data from earlier research \citep{dricker1996,Aleshin2019} (see Code Availability section for this data). 
Additionally, we adopted 1D profiles to construct the 3D seismic image in the arias of interest.  
We had a relatively small amount of input data. The spline interpolation like in \citep{horspool2006} or kriging technique as described in \citep{kozlovskaya2008} can not be considered optimal in this kind of problem. 
Instead, we preferred to use a machine-learning approach and chose the k-nearest neighbors (kNN) algorithm for its simplicity. 
One more reason to choose kNN is the data behind the analysis as it consists of separate sets of three different data types. 
The first one is the “yes-or-no” presence of the converted phase right after the $P$-wave arrival in the receiver function’s waveform. 
The second type is a series of $S$-wave velocity vs. depth pairs. 
The third type is crust thickness beneath the stations. 
Universally, the machine-learning approach allows us to process all of these types. 
In the “Data” and “Method” sections we described the input dataset and covered the processing algorithm features, correspondingly. 
We applied the method to the Moho depth data in the section “Moho Boundary Shape”. 
Notably, the results appeared to be in good alignment with previous works on the Moho shape of this particular region. 
In the next section, we applied the machine learning approach to locate and describe the structure of the LVSL regions. 
In the “Discussion” section we argue in favor of post-glacial rebound as the origin of the low $S$-velocity layer in central Finland.

\section{Data}

For our analysis, we predominantly utilized the results of the receiver function study of passive seismic experiments data, namely SVEKALAPKO and POLENET/LAPNET. Additionally, we used the results of the previous receiver function study for stations LVZ and APA \citep{dricker1996}; PITK and KEMI \citep{Aleshin2019}. Finally, we turned to account for the models derived from wide-angle seismic experiments \citep{janik2007,janik2009}.

\begin{figure}
\centering
\includegraphics[width=0.75\textwidth]{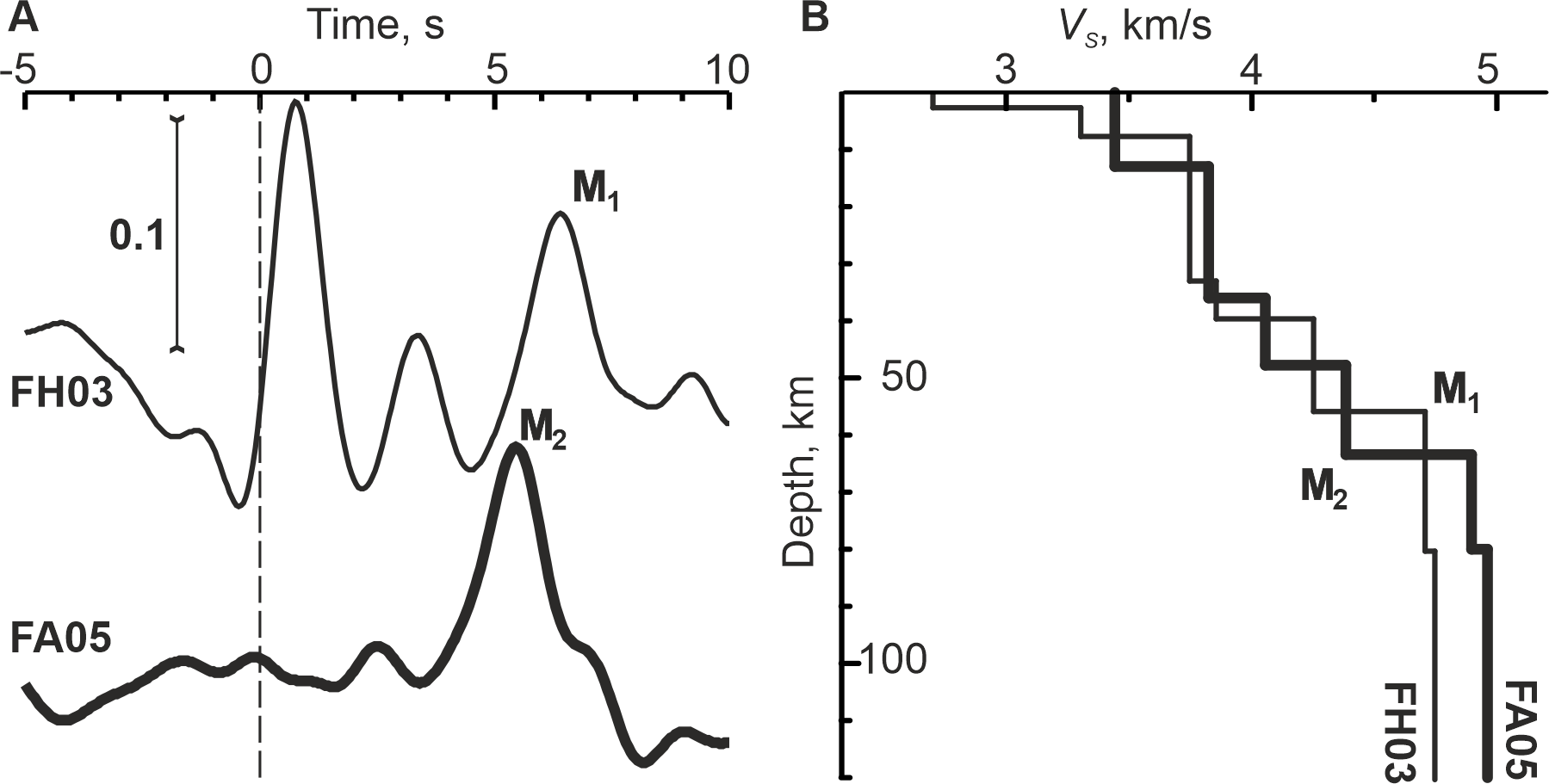}
\caption{A. Receiver function waveforms of two SVEKALAPKO stations, FH03 (thin line) and FA05 (thick line). FH03 waveform shows clear arrival of the converted phase near the first second after the $P$-wave arrival. This indicates the $S$-velocity jump immediately beneath the station. FA05 doesn’t show such a phase. The letters M$_1$ and M$_2$ pick to the Moho border phase. B. Corresponding 1D $S$-velocity profiles obtained through the receiver function inversion \citep{kozlovskaya2008}. The Moho borders are pointed with the letters M$_1$ and M$_2$.
}
\label{fig:depthvs}
\end{figure}
The LVSL under the station is manifested by the strong converted phase arriving within the first seconds after the main $P$-wave. 
Such a phase is clearly seen on the receiver function waveform (Figure \ref{fig:depthvs}, panel A, the FH03 station). 
We used it to construct the first dataset (dataset “A”) that represents the spatial distribution of the LVSL. 
Dataset members consist of the station name, location, and logical condition of whether the LVSL is present or not.  
A few additional dataset members were included from the models derived from wide-angle seismic experiments \citep{janik2007, janik2009}. It was not possible to draw a definite conclusion on the absence or existence of the LVSL under the stations FB11, FD07, LP11, FJ01, LP61, ARE0, LP75. 
These points were excluded from the dataset. As a result of the analysis, we obtained a set of points with the given values of the Boolean type. 
The dataset “A” enables us to compute the probability of the LVSL presence at an arbitrary point. 
To get more details on the structure we require a spatial distribution of elastic parameters. 
We have obtained it from the 1D $S$-velocity profiles by joint inversion of $P$-receiver functions and surface wave dispersion. 
Those profiles were adopted from \citep{kozlovskaya2008}. 
For the sake of convenience, we quantized the original layered models with 0.1 km step. 
The choice of the step is more or less volitional. 
The lower step limit is determined by computation time and complexity. The upper step boundary is limited to the thickness of the thinnest layer across all stations, which appeared to be 0.7 km for our case. 
Therefore, we’ve got a second dataset, “B”. 
It contains the station name, location (latitude, longitude? and depth), and the corresponding $S$-velocity. 
Members of the dataset are irregularly spatially distributed. 
1D $S$-velocity profile also provides for determining the Moho depth under the station (see fig. 1, panel B). 
POLENET/LAPNET \citep{silvennoinen2014} and SVEKALAPKO \citep{kozlovskaya2008} projects used the same approach to determine the Moho depth. Additionally, the Moho depth for stations LVZ, APA, and PETK, KEMI were adopted from \citep{dricker1996} and \citep{Aleshin2019}, respectively. Numerical values of the Moho depth for all the stations are unified into the third type dataset “C”. 
%The values are given in the scripts in supplementary. 
We used this dataset to construct the Moho depth map.

\section{Method}
At this point, we have three separate datasets with different types of data: two-dimensional logical values (dataset “A”), three-dimensional floating-point values (dataset “B”), and two-dimensional floating-point values (dataset “C”). 
We need to obtain the synthetic value for all types of data in datasets at an arbitrary location and depth, where applicable, we opted for using machine-learning methods to do this universally. 
To perform this task we have to, firstly, define the distance rules for each type of data. For data in dataset “A” and “C”, we can utilize the orthodrome (the geodetic line length). 
However, due to three-dimensional data in the dataset “B”, the cartesian distance is more convenient. 
Because the region of interest is quite small we switched from geographical latitude $\phi$ and longitude $\lambda$ to a local Cartesian coordinate system $(x,y)$ using approximate formulae

\begin{equation}
\label{eqn:cartes}
x = 2\pi R_E (\lambda - \lambda_1) \cos{\phi_C} / 360;
y = 2\pi R_E (\phi - \phi_1) / 360;
\phi_C = (\phi_2 - \phi_1) / 2;
\end{equation}
Here ${R_E}$ is the Earth’s radius, $\phi_1$, $\phi_2$ are the minimal and the maximal latitudinal boundaries of the region, correspondingly; $\lambda_1$ is the minimal longitude. Also, the station elevation was set to be zero as we did not include the terrain features into the consideration. There is no prominent influence of the simplifications on the conclusion; we used that due to convenience only. Thus, the distance between two points $\vec{r_i}=\{x_i,y_i,z_i\}$ and  $\vec{r_j}=\{x_j,y_j,z_j\}$ was defined as 

\begin{equation}
\label{eqn:distance}
 R(\eta; |\vec{r_i}-\vec{r_j}| = \sqrt{\eta^2(z_j-z_i)^2+(x_j-x_i)^2+(y_j-y_i)^2}.
\end{equation}

$Z$-axis is assumed normal to the horizontal plane $(x,y)$ plane and zeroes at sea level. The meaning of the scaling factor $\eta$ will be explained later.

As soon as we have defined the distance we should choose a formal mathematical model (algorithm) to be able to calculate the target value for any given location. 
In general, the algorithm contains a few inner tuning parameters. We can fix them with the input data (the algorithm learning). We’ve chosen the kNN-algorithm, one of the basic machine-learning algorithms. The kNN does not imply any preliminary training procedure. It is evaluated only to predict the target value $F$ at the given location $\vec{r}$ by the set of known values $\\{f_i\\}$ at given points $\vec{r_i}$

\begin{equation}
\label{eqn:target}
 F = F(\eta,K;\vec{r})=\sum_{i=1}^{K}w(\rho_i)f_i(\vec{r_i})/\sum_{j=1}^{K}w(\rho_j),
\rho_i = R(\eta; |\vec{r}-\vec{r_i}|).
\end{equation}
Here $w(\rho_i)$ is the weight function that depends on the distance from the current point $\vec{r}$ to the location of the $i$-th dataset member. The summation is performed over the $K$ dataset members closest to the point $\vec{r}$. Usually, the weight function is chosen in the form of a power law:

\begin{equation}
\label{eqn:powerlaw}
 w(\rho_i) ~ 1/\rho_i^\alpha, \alpha > 0.
\end{equation}

In this paper, we used this law with the value $\alpha = 1$. The {\it kNN} -algorithm is called a lazy learner because it implies no separate training process. In this algorithm, the dataset is stored in the memory in a certain spatial order based on the distance (\ref{eqn:distance}). Formula (\ref{eqn:target}) shows that inner parameters of the algorithm are just $K$ points of the input data nearest to a given position. The computational consumption originates from the search of the members closest to the given point. The complexity of the brute force selection from $N$ elements is about $O(N)$, which is quite inefficient. However, if the data is stored using the binary space partitioning (BSP) scheme the complexity can be reduced to $O(\log N)$. In our work, we used NumPy \citep{harris2020} arrays to store the data, which enabled hassle-free k-d tree-like BSP or ball-tree storage organization.

Along with inner parameters, the algorithm used in machine-learning often has hyperparameters, which do not change during training. In our case, these hyperparameters are the number of nearest neighbors $K$ and scale factor $\eta$ from (\ref{eqn:distance}). We chose to determine the optimal values of $K$ (and $\eta$ for dataset “B”) using the cross-validation technique \citep{hastie2009}. The main idea is to split the input data into training and validation blocks. Training blocks are used to fit the internal parameters of the model. Then, the trained model is used to predict the data of the validation block. The comparison of the predicted data and the validation data is used to estimate the training quality. 

Within the actual implementation, the initial set is to split into $p$ blocks, which we will call folds, each of the same size. Each fold is consistently served as a validation block, and the remaining $p - 1$ folds act as the training blocks. The main advantage of this strategy is that each piece of data is used both for training and for quality control. There are a lot of strategies for splitting data. For dataset B we chose $p = 5$ (see below). In problems with a lack of data, that is the case for datasets “A” and “C”, the value of $p$ is appropriate to be chosen equal to the number of members of the source dataset so that each fold contains just a single point. This strategy is called Leave-on-Out.
To estimate the training quality, we have to define the quality function. In regression analysis one of the simplest choices is a mean squared difference between the values calculated from the training block and the corresponding samples in the validation block. Essentially, this metric is a measure of the mean interpolation error. We have chosen such a quality function

\begin{equation}
\label{eqn:hyperp}
\mu(K, \eta) = \sqrt{(1/N)\sum_{p=1}^{P}\sum_{l=1}^{L}(F(\eta,K;\vec{r_l})-f_l)^2} 
\end{equation}
to determine hyperparameters $K$ or $K$, $\eta$ in the Moho boundary shape (dataset “C”) and $S$-velocity 3D-model problems (dataset “B”), correspondingly.
However, this metric is unsuitable for the calculation of areas covered with a low $S$-velocity layer (dataset “A”). For this problem, we have to formulate the task as a binary classification problem. Within this classification, we divide all the points into two classes. Let the point belong to class 1 if the low $S$-velocity layer at the location of the point exists and to class 0 otherwise. Therefore, our goal is to compute the probabilities that a given point on the surface belongs to class 1 or class 0. In this setting, receiver operating characteristic (ROC) is commonly employed. The ROC curve is created by plotting the true positive rate (probability of detection) versus the false positive rate (probability of false alarm) generated by the binary classifier. The area under the curve (AUC) 	gives a measure of the classifier quality (more detail can be read elsewhere, \citep{fawcett2006}).The rest of the calculation details are the dataset specific and will be described later.

\section{Moho Boundary Shape}
\begin{figure}
\centering
\includegraphics[width=0.75\textwidth]{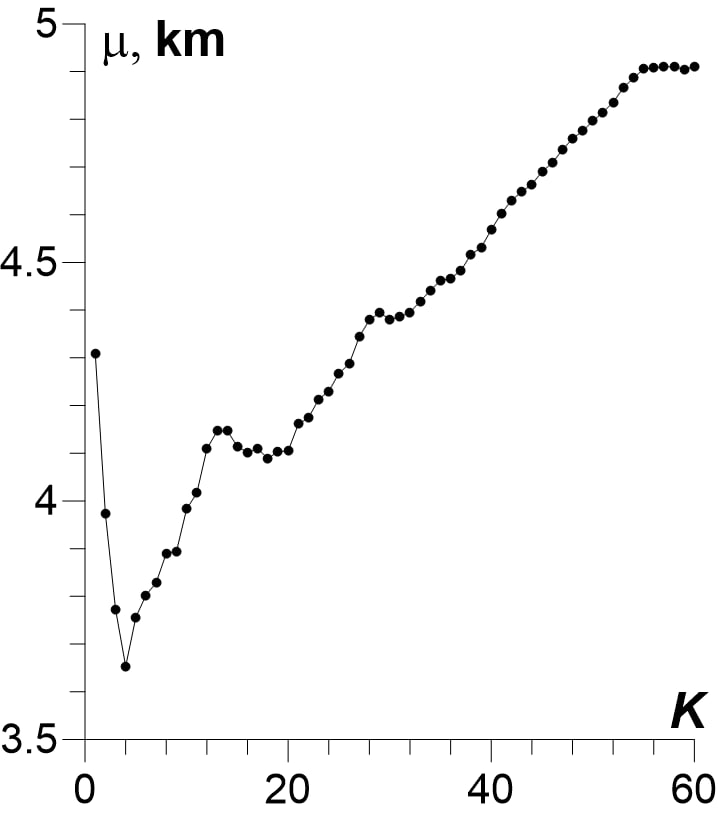}
\caption{The plot of the quality function (\ref{eqn:hyperp}) versus the number of nearest neighbors $K$ derived from the Moho boundary depth dataset.}
\label{fig:kmu}
\end{figure}

For this region, the depth to the seismic Mohorovicic discontinuity approximately equals the compositional crust-mantle boundary. The shape of the Moho boundary with receiver function technique was previously studied for several areas in the region: Kola in \cite{kosarev1987}, Southern Finland in \cite{alinaghi2003,kozlovskaya2008,grad2012},  Northern Finland in \cite{silvennoinen2014}. In a series of works the Moho was investigated with controlled source seismic experiments \cite{grad1992,janik2007,uski2011}. In \cite{grad2012,uski2011} the crust thickness was studied as part of the European lithosphere structure. In the paper by \cite{silvennoinen2014}, the topography of the Moho under the northern part of Finland was drawn by a joint interpretation of the receiver function and controlled source experiments. In our paper, however, we only use the data obtained with the receiver function technique. We opted for this limitation because as shown in \cite{silvennoinen2014} the Moho boundary estimated from the receiver function may differ from that of a controlled source experiment.
Several reasons can explain this difference.  In the controlled source approach the Moho depth is obtained by $P$-waves in most cases, while the receiver function utilizes $P$- or $S$-waves from teleseismic sources converted at the Moho. Also, the frequency content of the signal from teleseismic source data is rather distinct from that in near-vertical reflection, wide‐angle reflection, and refraction experiments. Furthermore, rays in both the near-vertical setup and the receiver function are sub-vertical apart from the wide-angle approach. The other reasons are region-specific. In the southern part of the region, the lowermost part of the crust has a high-velocity layer \citep{janik2007} that masks the high-velocity contrast at the Moho for $P$-wave \citep{janik2009}. Also, there is a tangible seismic anisotropy in the upper mantle \citep{vecsey2007,VINNIK2016,yanovskaya2019} that causes the discrepancy in sub-lateral and sub-vertical $P$-wave travel times.
The number of nearest neighbors $K$ is only one hyperparameter in the Moho boundary shape problem. To determine an optimal value of $K$ we adopted the Leave-one-Out cross-validation procedure. We used the quality function (\ref{eqn:hyperp}) (we set $\eta$ = 0). The optimal value of $K=4$ minimizes the functional (see fig. \ref{fig:kmu}). The corresponding mean interpolation error is 3.7 km.
\begin{figure}
\centering
\includegraphics[width=0.75\textwidth]{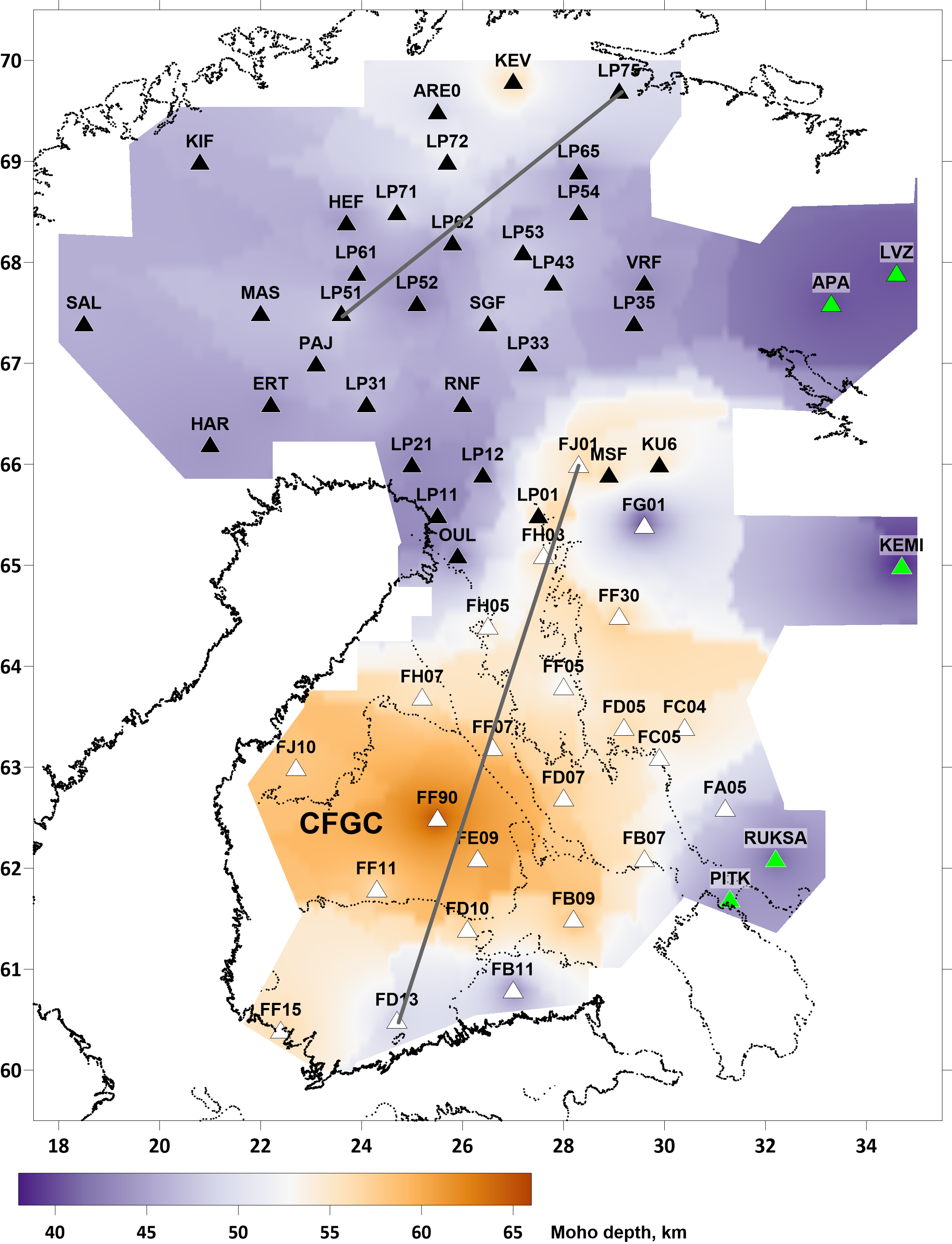}
\caption{Depth to the Moho boundary. Colored triangles mark the positions of the input points. The color of the symbols corresponds to the source used: black — \cite{silvennoinen2014}; white — \cite{kozlovskaya2008}, green — \cite{dricker1996,aleshin2006,Aleshin2019}. The abbreviation “CFGC” means Central Finnish Granitoid Complex.}
\label{fig:mohodepth}
\end{figure}
The map in Fig. \ref{fig:mohodepth} depicts the depth to the Moho boundary, red —-- data from this study. The map was constructed with formulas (\ref{eqn:cartes}),(\ref{eqn:distance}),(\ref{eqn:target}) where $K=4$. The maximum thickness of the crust appears to be in the center of the southern part of the region which corresponds to the results described in \cite{alinaghi2003,kozlovskaya2008,silvennoinen2014}. Additionally, there is a correlation of the Moho boundary with the Central Finnish Granitoid Complex shapes, which is also discussed in cited papers. 
\begin{figure}
\centering
\includegraphics[width=0.75\textwidth]{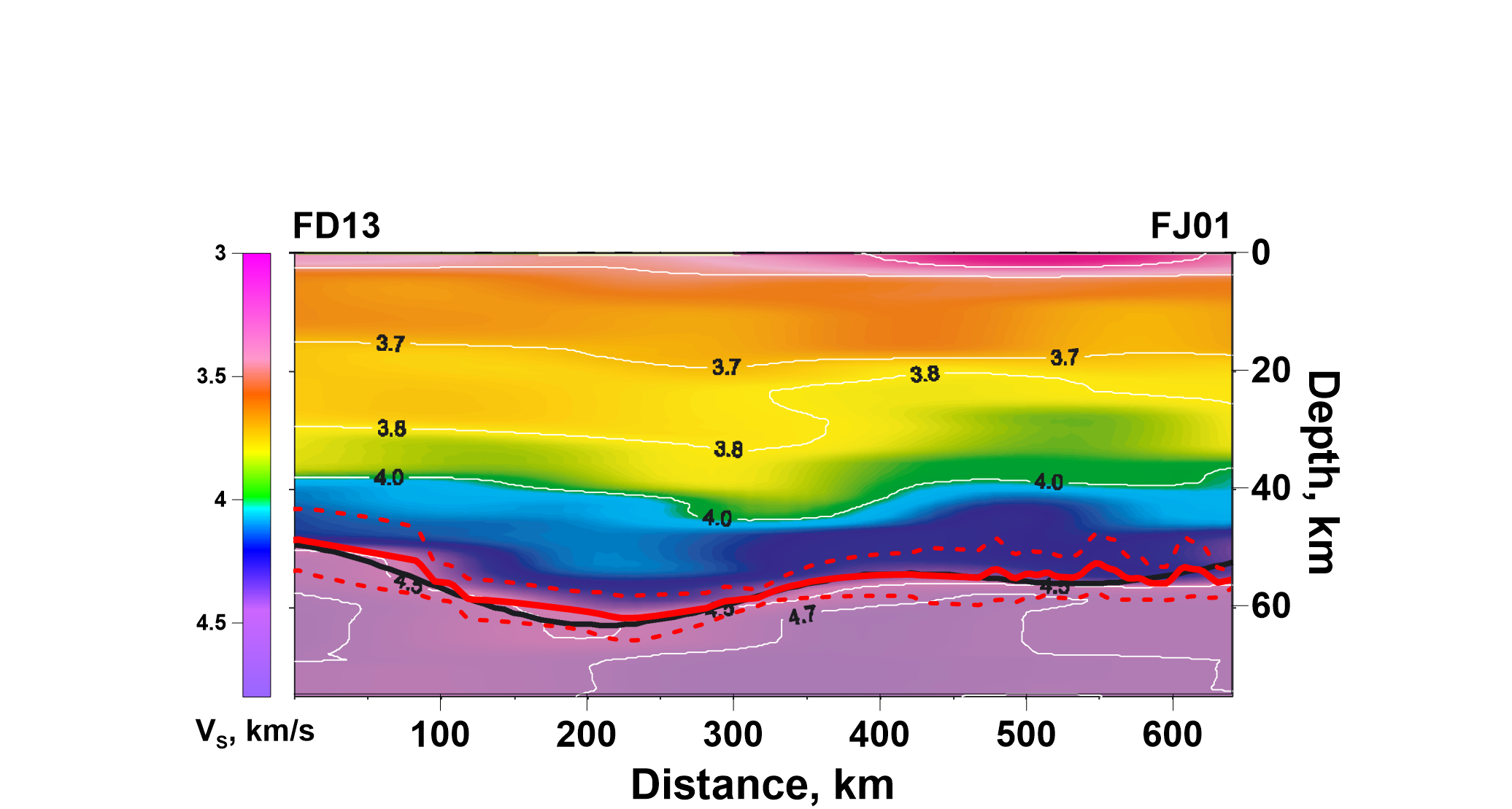}
\caption{ Depth to Moho along with the FD13-FJ01 profile. Black and red lines mark a cross-section of the Moho shape obtained with the kriging technique and the machine learning {\it kNN} method, correspondingly; red dashed lines show the {\it kNN} interpolation errors. The line oscillations at distances of 550-600 km are likely due to erratic data for station FG01.}
\label{fig:fd13fj01}
\end{figure}
Fig. \ref{fig:fd13fj01} shows the $V_S$-profile along FD13-FJ01. The black line marks the Moho depth obtained in \cite{kozlovskaya2008} with the kriging technique. The solid red line denotes the Moho boundary obtained in this study with the {\it kNN}  approach. Both lines agree well, showing good consistency among the approaches.  Dashed red lines show interpolation error intervals. The map (Fig. \ref{fig:mohodepth}) and the profile (Fig. \ref{fig:fd13fj01}) reveal the crust thickness anomaly under station FG01. The same was noted by \cite{kozlovskaya2008}. It is worth noting that interpolation misfit surges in the vicinity of FG01, hitting values times the mean error. Thus, we lack the confidence to conclude on the Moho depth there.
\section{Map of the low $S$-velocity Layer in the uppermost crust}
\begin{figure}
\centering
\includegraphics[width=0.75\textwidth]{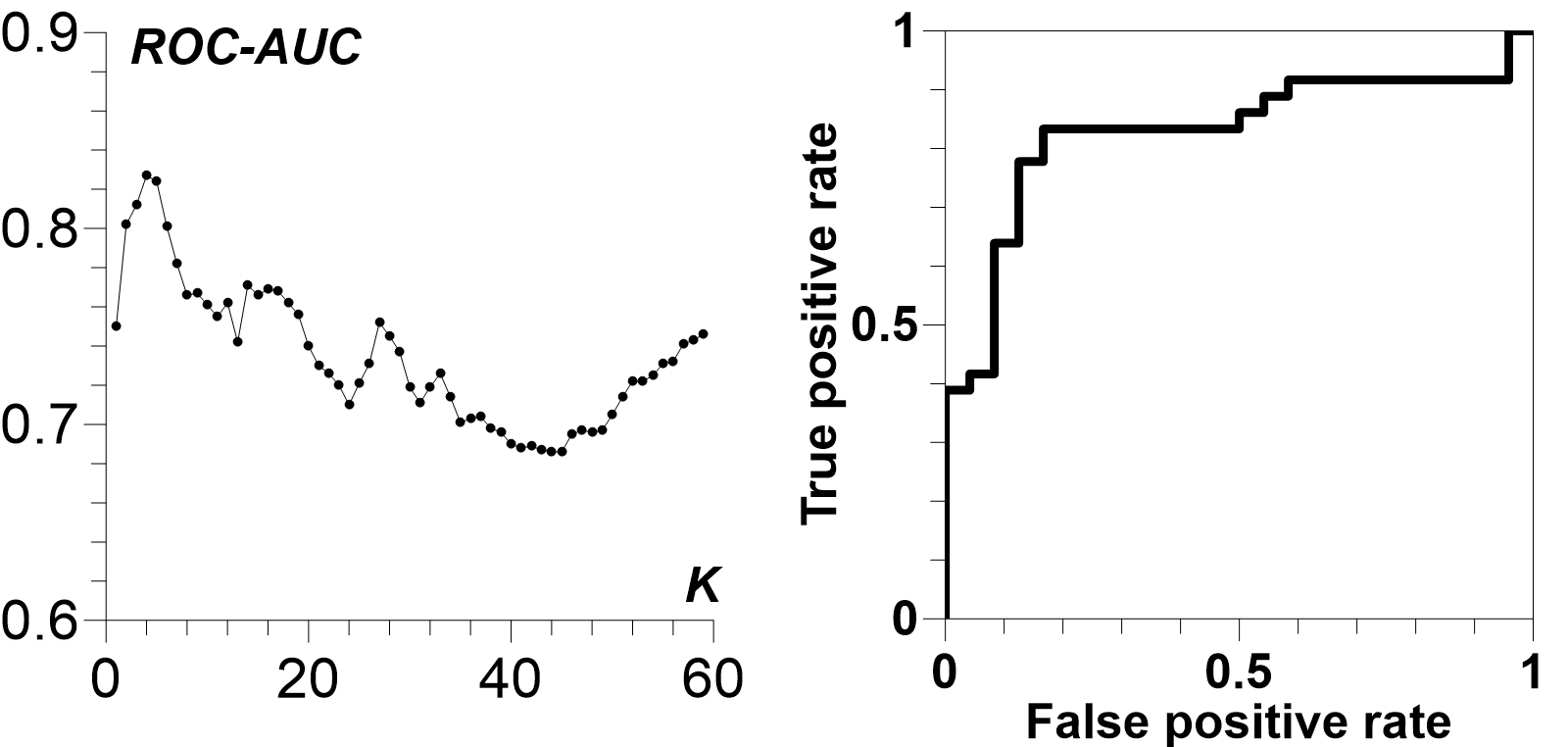}
\caption{The plot of the ROC-AUC versus the number of nearest neighbors $K$ (left). The plot of ROC for the optimal value of $K=4$ (right).}
\label{fig:rocauc}
\end{figure}

Now, we can apply the machine learning approach to dataset “A” to build a map of the LVSL coverage in the region. At first, we have to determine the optimal number of nearest neighbors $K$. We used the ROC-AUC metric and applied the Leave-one-Out strategy for cross-validation, similar to the Moho shape calculation. The larger ROC-AUC value corresponds to the better classification model. From the left panel of \ref{fig:rocauc}, one can see that the optimal number of nearest neighbors is reached at $K=4$. The shape of ROC at this value is plotted on the right panel.
To construct the map of LVSL from the dataset “A” we substituted $\eta=0$ and $K=4$ in formulae  \ref{eqn:cartes}, \ref{eqn:distance}. The calculation was performed at nodes $\vec{r_n}$ of the regular surface grid. By the definition, every member of the input dataset  “A” is either 0 or 1. It leads to the function $F_n=F(K,\vec{r_n})$ from formula (\ref{eqn:hyperp}) is constrained to $ 0 \leq F_\alpha \leq 1 $. Thus $F_n$ might be treated as the probability of the presence of the LVSL at the given point $\vec{r_n}$. To perform the final classification we had chosen two threshold values $t_0$ and $t_1$. Whether the $F_n$ was less than $t_0=0.4$ we concluded that there was no LVSL at the point $\vec{r_n}$. As a result, the point was classified as class “0”. Otherwise, when $F_n$ exceeded $t_1=0.6$ we decided that there was an LVSL at this pint, and $\vec{r}_n$ was classified as class “1”. The interim values of $F_n$ that comply with $(t_0<F_n<t_1)$ remained unclassified and formed the “buffer zones” between two classes. 

\begin{figure}
\centering
\includegraphics[width=0.75\textwidth]{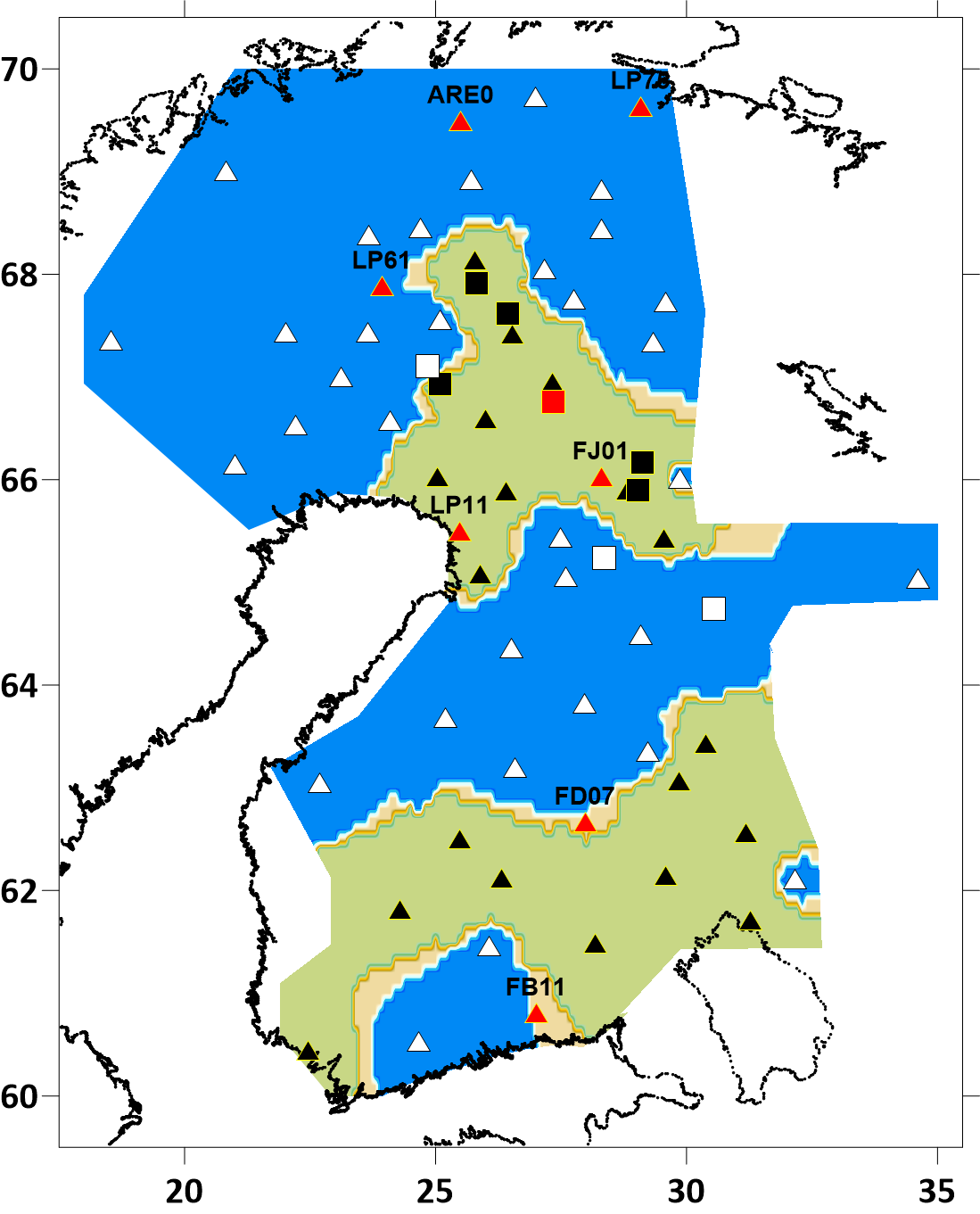}
\caption{Map of the low-velocity surface layer calculated from the dataset “A”.  Areas in which the probability of the presence of the LVSL exceeds 60\% are colored in blue. Olive shaded areas denote the probability is less than 40\%. Triangles indicate the position of seismic stations and squares indicate the points from wide-angle experiments. They are white or black, depending on the presence or absence of the LVSL at the point, respectively. Initially excluded points are marked red.}
\label{fig:datasetAmap}
\end{figure}

The map obtained is shown in Fig. \ref{fig:datasetAmap}. The map contains three separate LVSL areas in the northern, central, and southern parts of Finland. Stations excluded from the analysis are mostly located near the borders of those areas like FB11, FD07, LP61, or near the outer border of the region like ARE0, LP75. Such distribution explains partly the difficulties of the initial classification. With new results, we can suspect the existence of the LVSL under the station LP61. There appears to be no such layer under the stations LP11, FJ01, and in point HT. The stations FD07 and FB11 are located inside the “buffer zones” so we can not decide the LVSL under these points.
The southern and northern LVSL areas could be linked to the geology: the metamorphosis of bedrock and the presence of Paleozoic sedimentary rocks atop the archean basement, respectively. In the central part of Finland, the LSLV can be explained with neither the composition nor age of the rocks. As seen in Fig. \ref{fig:datasetAmap} the LVSL area in central Finland covers both the Archean and Proterozoic rocks. The origin of the layer should be discussed separately. But before that, we used dataset “B” to highlight the structure under this area. 
\begin{figure}
\centering
\includegraphics[width=0.75\textwidth]{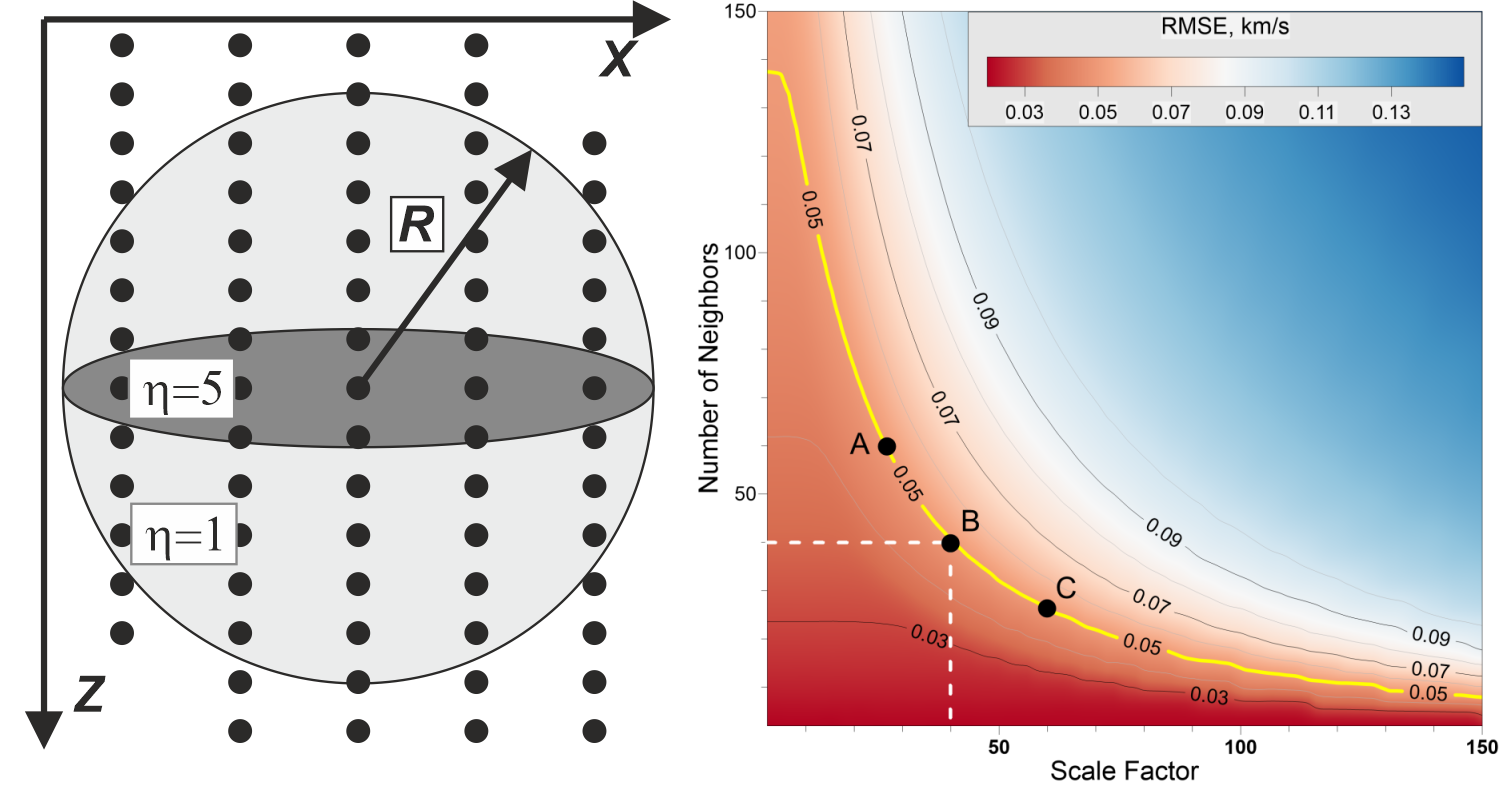}
\caption{Left: Illustration of the role of scale factor. Right: plot of root-mean-square deviation (formula (\ref{eqn:hyperp}))  for the different values of $K$ and $\eta$.}
\label{fig:nnScale}
\end{figure}
We have to mention that when we quantized the 1D $S$-velocity profiles we got vertical distances between the points of the same profile much smaller than distances between the stations. 
Here, to illustrate this issue we would stick with the alternative formulation of kNN where the term “nearest neighbors” assumes all the points inside the sphere of radius R around the given position, see Fig. \ref{fig:nnScale}, left. This corresponds to set $\eta=1$ in the formula (\ref{eqn:distance}). With moderate R the nearest points would originate from the same profile. The issue could be solved by using the ellipsoid instead of the sphere and setting $\eta>1$. We used a fixed the nearest neighbor number $K$  instead of the radius of the sphere $R$ and the factor  controls the $Z$-axis scaling.
To determine the optimal $K$ and  $\eta$ we used the cross-validation technique for dataset “C”. The dataset “C” has quite a large number of points so we’ve split it into 5 folds of approximately 8000 points each. The quality function $\mu(K,\eta)$ calculated for $K=1,2,...,150$ and $\eta=1,2,...,150$ is shown on Fig. \ref{fig:nnScale}, right. Such a distribution is intrinsic to concurrent optimization problems, as seen in, for example,  \cite{chiu2009}. Optimal pairs of  $K$ and $\eta$ are located where the function $\mu(K,\eta)$ is minimal. These values of $\mu$ are placed at the axes origin on Fig. \ref{fig:nnScale}, right. However, this isn’t the case in our problem because picking such low values will lead to very low K and $\eta$ and subsequently to an overinterpretation. Larger values of $K$ and $\eta$ correspond to smoother interpolation results, and to larger $\mu(K,\eta)$. Meanwhile, the error of interpretation cannot exceed the error of the receiver function method, which is typically about 0.1 km/s. Overcauteously, we chose 0.05 km/s for the $\mu(K,\eta)$. This level provides for the optimal pairs of $K$ and $\eta$. For example, picking points A, B, and C from Fig. \ref{fig:nnScale}, right net visually equivalent results. So we picked point B, which is the closest to the origin. Such a choice is obvious for multi-objective optimization problems \cite{chiu2009}. We visualized the horizontal section of the $S$-wave velocity at the surface level ($z=0$), see Fig. \ref{fig:vsmap}, left. The low velocity region matches the probabilities computed with dataset “A” well. There are strong $S$-velocity variations of the LVSL in archean, proterozoic, and transition regions. We can speculate that the thickness of the LVSL could depend on rock composition, however, investigation of this dependency is out of the scope of the current work. We also imply that the $S$-velocity depends on the amount of rock fracturing and water saturation. 
\begin{figure}
\centering
\includegraphics[width=0.75\textwidth]{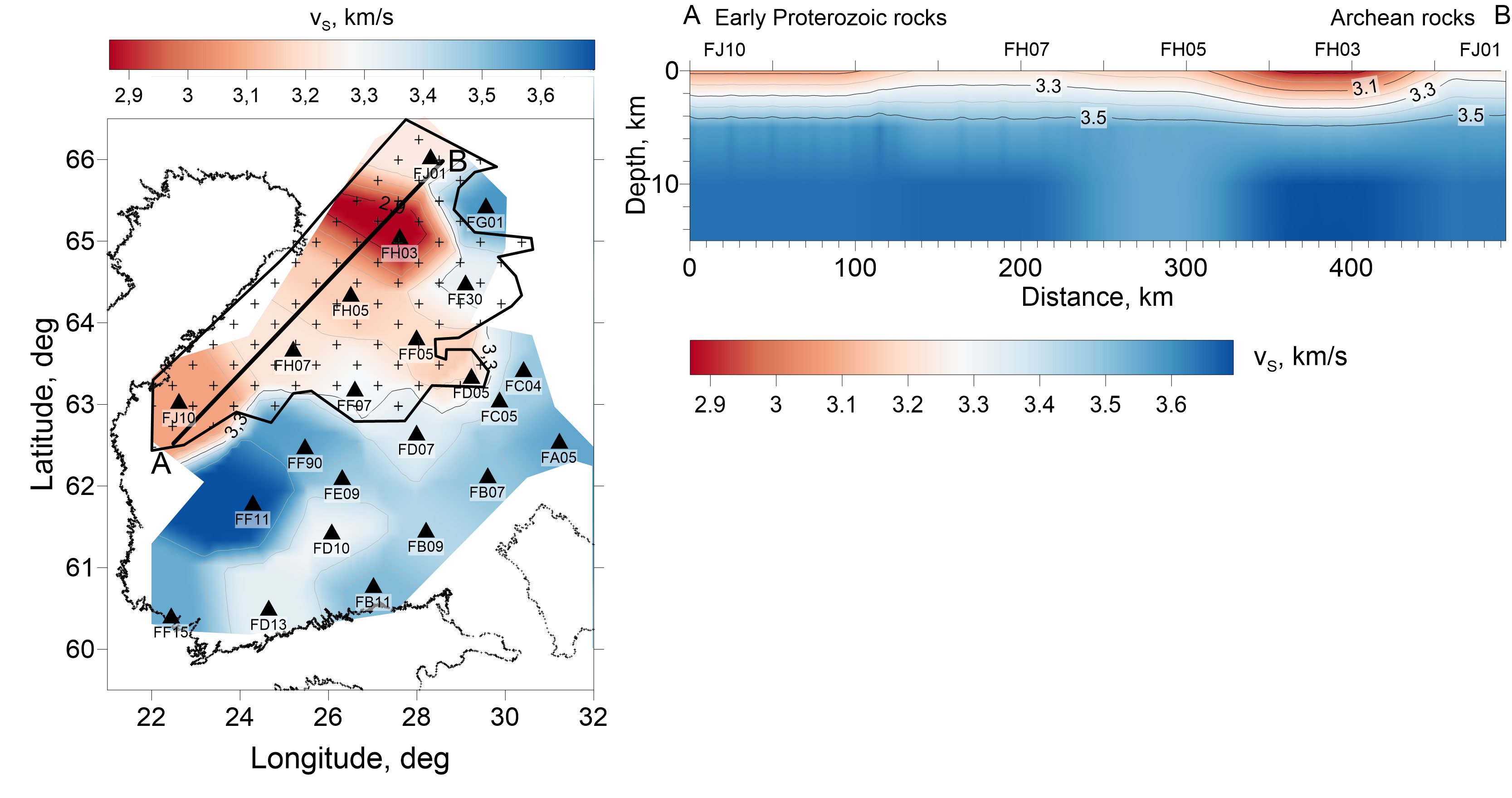}
\caption{ Left: Velocity map of $S$-waves at the surface. Plus-marked (“+”) area matches the central LVSL-area on Fig. \ref{fig:datasetAmap}. Right: vertical section of the $S$-wave velocity of AB profile.}   
\label{fig:vsmap}
\end{figure}
\section{Discussion}
In our study, we do not make a particular interpretation of $S$-wave velocity values of the low $S$-velocity layer in the uppermost crust. The most interesting fact is that this layer has been revealed both in Archean and Proterozoic areas and it is existing in bedrock units with different compositions. Moreover, it is not correlating with the rock units composed of low-velocity rocks, like granites. For example, the layer is not detected in the Central Finland Granitoid Complex. Therefore, its origin seems to be not dependent on the tectonic age or composition of the bedrock and rock fracturing is the main factor responsible for low velocities.  On the other hand, this layer is not distributed uniformly around the Fennoscandia and it is present in several large-scale areas indicated by dark blue in Fig. \ref{fig:vsmap}.  These two features (e.g. absence of correlation with tectonic age and rock composition and systematic distribution in large-scale areas) need to be explained.
In our study, we used the converted waves with dominating periods of 2 s, which means that the wave is averaging fractures of different scales up to 1 km. Such large-scale fractures reaching the depth of 1 km could not form due to rock weathering, but rather due to some large-scale processes.  One of the explanations of this layer could be that it has been formed due to the processes related to multiple episodes of deglaciation of the Fennoscandian Shield. 
Relation between intraplate seismicity and processes of deglaciation (Post-Glacial Rebound, PGR) has been studied previously by numerous authors, see \cite{turcotte2002,lagerback2008,slunga1991}. The seismicity at the ice sheet margins is attributed to unloading processes and to ice melting (\cite{wu2000}). According to \cite{spada1991}, the PGR is the valid reasoning for normal and lateral crustal deformation at polar regions. The PGR appeared to have a solid role in stress redistribution. [Steffen et al., 2012; Steffen et al., 2014] unfold that in the cases of Greenland and Antarctica the earthquakes mainly credit to the PGR. However, the previous studies of seismicity in the Fennoscandia \citep{lagerback2008, slunga1991, uski2003, uski2012} were mainly trying to explain the origin of moderate-to-large magnitude earthquakes (M>3) with sources corresponding to large-scale faults. Studying the effect of the PGR on the formation of smaller scale fractures in the bedrock has not been done previously. 

Nowadays climate change caused rapid warming and melting of ice sheets of Greenland and Antarctica.  Before this massive melting, these areas were considered as generally aseismic, which has been explained by the suppressing of earthquakes due to ice sheet load. The rapid deglaciation results in significant stress change in these areas, both in the crust and in the mantle, and hence in new seismic events in these areas. That is why more detailed studies of the connection between PGR and bedrock fracturing at different scales are possible nowadays. In particular, \cite{olivieri2015} studied the relationship between present-day crustal deformations and spatio-temporal distribution of recent earthquakes in Greenland. They showed that nowadays, Greenland undergoes the significant deformation due to two processes: the melting of the Greenland Ice Sheet  (the “elastic rebound”) and the PGR from late-Pleistocene. So, \cite{olivieri2015} suggested two distinct mechanisms responsible for seismicity: regional crustal earthquakes are attributed to the PGR, and local events are attributed to the ER and present-day ice melting. In the areas with the largest ice sheet thickness, the seismicity is suppressed nowadays. If we consider deglaciation of Greenland as a modern analog of processes of Fennoscandia deglaciation, we may speculate that the LVSL in the Fennoscandia was formed due to ER in the areas which underwent rapid deglaciation in the past. 
Previous studies of Fennoscandian ice sheet melting and retreat models and reconstructions have been summarized by \cite{stroeven2016}.  They showed that the deglaciation rate was significantly different across the area investigated in our study. These studies are based on geomorphological and geochronological data. However, the amount of these data is limited, which increases the uncertainty in the modeling of ice sheet deglaciation processes. We think that our study provides new, previously not considered evidence of the effect of deglaciation on the fracturing of the uppermost bedrock that could be used both for upgrading of Fennoscandia deglaciation models and in predicting the consequences of climate change and melting of Greenland and  Antarctica Ice Sheets.  
\section{Conclusions}
The results obtained continue and generalize the series of studies conducted by the authors. This time, we performed a joint interpretation of the results of previous studies supplemented by data for several new stations using a modern mathematical apparatus. The results showed the effectiveness of universal machine learning methods for analysis and generalization. Some algorithms of machine learning benefit highly from a data deficit that is typical in many geophysical studies. Further progress in exploring the region is hindered by the lack of data on a significant part of the Russian territory.

\section{Acknowledgments}

The majority of this work would not be possible without our late colleague Dr. Grigory Kosarev. 

\newpage

\textbf{Code Availability}
The source code and data are available for downloading at the link:
\url{https://github.com/iperas/paper-fennoscandia-ml}

\bibliographystyle{cas-model2-names}
\bibliography{bibliography} 

\end{document}